\documentclass[10pt,conference,letterpaper]{IEEEtran}
\usepackage{wrapfig}
\usepackage{graphicx}
\usepackage{algorithmic}
\usepackage{algorithm}
\usepackage{epsfig}
\usepackage{amssymb}
\usepackage{subfigure}

\ifCLASSINFOpdf
\else
\fi

\begin{document}
\title{Empowering Evolving Social Network Users\\ with Privacy Rights}

\author{\IEEEauthorblockN{Hasan M Jamil}
\IEEEauthorblockA{Department of Computer Science\\
University of Idaho, USA\\
{\tt jamil@uidaho.edu}}
}
\maketitle

\begin{abstract}
Considerable concerns exist over privacy on social
networks, and huge debates persist about how to extend the artifacts
users need to effectively protect their rights to
privacy. While many interesting ideas have been proposed, no single
approach appears to be comprehensive enough to be the front runner. In this paper, we propose a
comprehensive and novel reference conceptual model for privacy in constantly evolving social networks and establish its novelty by briefly contrasting it with contemporary research. We also present the contours of a possible query language that we can develop with desirable features in light of the reference model, and refer to a new query language,  {\em PiQL}, developed on the basis of this model that aims to support user driven privacy policy authoring and enforcement. The strength of our model is that such extensions are now possible by developing appropriate linguistic constructs as part of  query languages such as SQL, as demonstrated in PiQL.
\end{abstract}

\begin{IEEEkeywords}
privacy; social network, evolving networks; content visibility; peer rules; exceptions; global accessibility;
\end{IEEEkeywords}

\IEEEpeerreviewmaketitle

\section{Introduction}
\label{introduction}

In its basic form, privacy means unwanted disclosure prevention of information content of an individual or entity\footnote{One concomitant issue is security that deals with threats of intrusion that breaks down the infrastructure put in place to protect the privacy in order to gain access to protected information. This paper is primarily focused on the privacy, and we take the view that security is an orthogonal issue that can be dealt with separately, and hence, we will not be discussed any further.}. As the number of individuals or the size of their information holdings in an information system grow, the complexity of privacy assurance also increases. This is mainly because the rules for visibility and accessibility grow more complex, and conflicts and exception become dominant making enforcement more difficult. In the context of social networks, the sheer number of users make it impossible for system administrators to develop privacy rules based on roles or rights. The responsibility is left to individuals or users to define their own privacy rules. However, it becomes a daunting task for users to individually define privacy rules for such an overwhelming number of users most of whom they don't even know. To further complicate matters, social networks are increasingly being exploited for commercial means which necessitates grouping the user base around various business objective functions that are often volatile, forcing frequent restructuring of the relationship network among the users. A comprehensive privacy model that accommodates all these competing needs and captures all these complex characteristics is still outstanding. The overall goal of this paper is to propose a reference conceptual model for privacy in evolving social networks that we believe elucidates a set of characteristics a good comprehensive privacy model should have, and discuss how the current trends fare in light of this model.

Privacy in social networks such as FaceBook, LinkedIn, Flickr, Twitter, Ning, and MySpace has been approached from multiple viewpoints and interesting solutions have been proposed. The solutions proposed thus far favor one or the other groups of users and fail to address many critical needs. Given the sheer size of the user base in these social networking sites and the diversity of the growing digital communities, it is safe to anticipate that the issue of privacy will remain active for the foreseeable future. As we will discuss in this paper, the issue of privacy permeates across many different aspects of social network infrastructure and due considerations must be given to each of them.

\subsection{Related Research}

Privacy on the internet, especially on social networks, has different meanings for different interest groups. The basic ingredients of a social network are a set of users, the ``actors", a set of information contents owned by these actors, their relationship with other users, and the behavior captured as actions they perform. These actions may involve sharing their information contents with other users they call friends, communicate with each other, express opinion, and so on. Such actions can be anything that people usually do in their physical world that can be transmitted electronically with some of these actions kept private, just as they do in the real world. The real value of these communities and the networks, however, is actually in the behavior a user exhibits as a member of the network they are part of, and not necessarily in the information content they own.

There is less debate about what is private content that each user owns \cite{Rosenblum07,BrandtzaegLS10}. Generally, all content of a user is considered private when it comes to other users, except the system and the network administrator \cite{Marsoof11,AhnSS11,GehrkeLP11}. There is some debate though about what privacy rights a user retains on the information he chooses to share by making it visible with a set of friends \cite{SquicciariniSW10,Schneier10a}, and whether he can revoke those visibilities at will \cite{NagleS09,SunZF10}. Or, does sharing mean a user gains the right to do ``whatever" with the information content she can access in the network \cite{AbdessalemD11}. The major debate, however, is over the use of the behavioral information of users that administrators can gather and use for their own purpose \cite{KahlCTR11,AdnanNKTRR11,Chou10,ZhaoLLC09,LiBS11,YeungI11}. The issue in point is whether they can do so without violating the basic privacy rights users would like to enjoy \cite{BhagatCKS10,HeVSAA09}.

\subsection{Our Contributions}

As mentioned at the outset, our goal in this paper is to propose a reference data model for social network to support a user defined privacy articulation mechanism and query language for fine-grain privacy policies. The proposed model in section \ref{model} is based on the results of a brief survey of contemporary research which is at the heart of this presentation. Although there are many interesting dimensions that can be used to compare existing models, in the interest of conciseness and focus, we will evaluate our model on five different axes. In section \ref{default}, we discuss issues related to default privacy rules. Privacy implications of dynamic networks is discussed in section \ref{dynamic}. User power to manage and control privacy rights is discussed in section \ref{user}. Finally, privacy models and their representation issues are presented respectively in sections \ref{theory} and \ref{language}. We hope to highlight the simplicity and potentials of the proposed reference model on intuitive grounds and discuss research that address the issues we raise. We believe scattered machineries are available to realize our model and develop a query language using it. We discuss the outline of a possible query language in section \ref{outline} that is capable of addressing the issues raised in a comprehensive manner before we conclude in section \ref{conclusion}. As an aside, we note that to substantiate our claim that development of a language to address these issues is quite possible, we refer readers to a new query language, called \underline{p}r\underline{i}vacy \underline{q}uery \underline{l}anguage or {\em PiQL} \cite{PiQL-PODS-2012}, that has been proposed recently based on the reference model introduced in this paper.

\section{A Conceptual Model for Privacy\\ in Evolving Social Networks}
\label{model}

\begin{figure}[h]
\centerline{\epsfig{figure=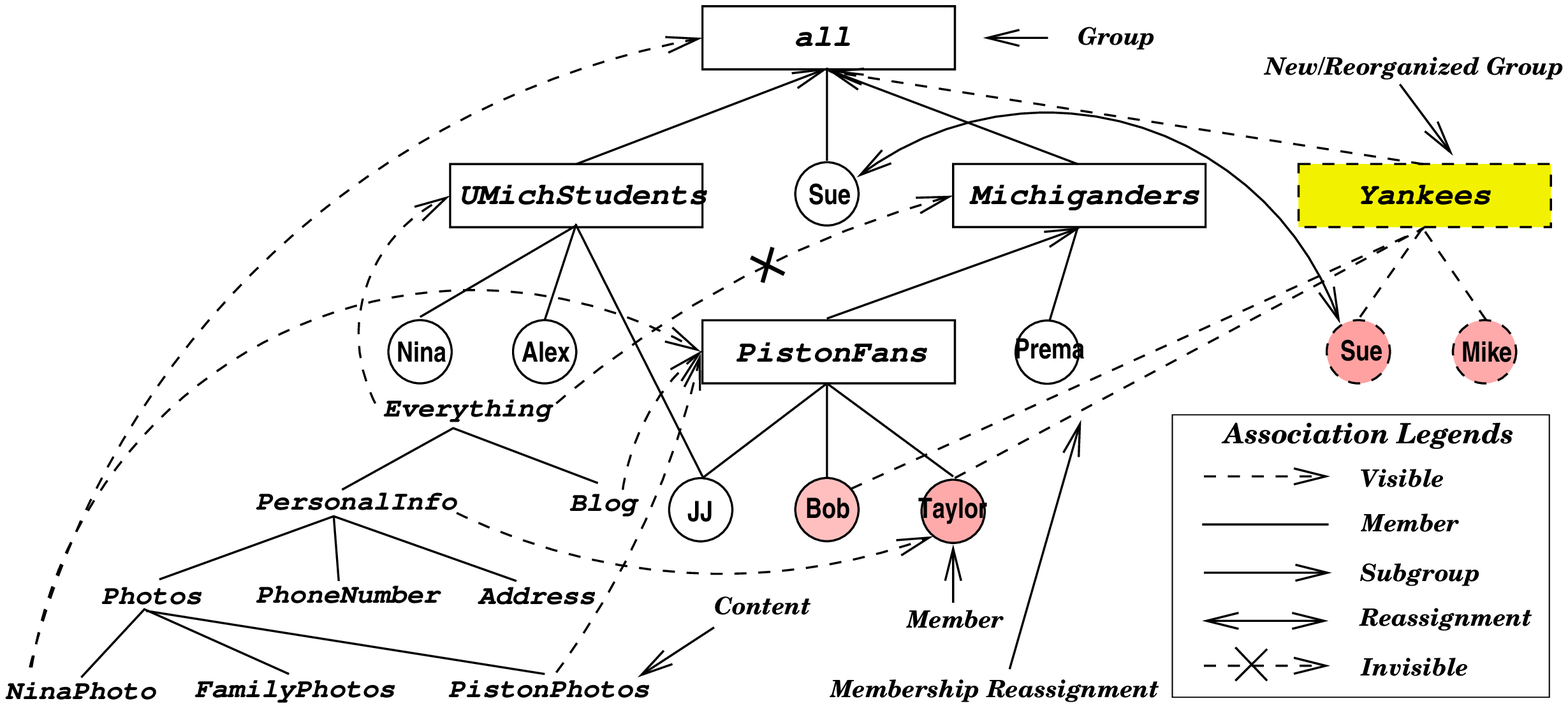,height=1.5in,width=.45\textwidth}}
\caption{Nina's content visibility on social net.} \label{hie}
\end{figure}

We introduce our conceptual network privacy model using the example shown in figure \ref{hie} on intuitive grounds. The model includes four main objects -- {\em members, groups, contents} and {\em associations}. Figure \ref{hie} shows a small hypothetical social network.

\subsection{Members, Groups and Contents}
\label{groups}

In this figure, members are shown as labeled circles, and groups are shown as labeled rectangles. Members are assigned to groups based on some arbitrary principles such as shared properties, membership of an institution, interests or location. For example, {\em Nina} and {\em Alex} are in {\em UMichStudents} group because they are UMich students. Members may belong to multiple groups as needed. These groups are dynamic and may be reorganized at any time without restrictions. Each member may have contents they own such as personal information, photos, documents, etc. that they introduce into the network or remove from it from time to time. In this figure, {\em NinaPhoto, FamilyPhotos}, and {\em Blog} are some of the contents {\em Nina} owns.

\subsection{Associations}
\label{hierarchy}

In this model associations are the vital tools  that make the network meaningful. There are three types of associations -- group and content hierarchies, and visibility assignments. Groups are organized in a directed acyclic graph in which a member in a group $G$ is considered a member of another group $S$ as well if $G$ is a subgroup of $S$. In this figure, {\em PistonFans} is a subgroup of {\em Michiganders}, and hence {\em JJ} is also a member of {\em Michiganders}. Since groups and members can belong to multiple groups as members of subgroups, {\em JJ} is shown to belong to two groups -- {\em UMichStudents} and {\em PistonFans}. Contents can be organized in a tree-like hierarchy\footnote{However, in this paper, we only consider the case where contents are organized in like tree structures to align ourselves with the directory structures of file systems. This design choice can be altered to a more general DAG like organization with no disruption to the proposed model. But we do not foresee any practical benefit to this alternative choice.}. In this figure, {\em Nina} organized her contents under {\em Everything} as a rooted tree.

\subsubsection{Content Visibility}
\label{content-vis}

Content visibility assignments are created by members and altered anytime without restriction. In this paper, we consider only two modes of visibility or access rights -- visible and invisible\footnote{We will discuss, in a later section, how this binary visibility model can be extended to include a lattice like access right policy to include read, write, update, and republish type of  authorizations.}. Visibility of any node in the members' content hierarchy in one of the two modes is assigned to any node in the social network. The logical implication of such an assignment follows the notions of inheritance in object-oriented systems with overriding. Hence, a visibility mode assigned to a network hierarchy node remains implicitly in effect for the subgroups and members of the group until it is overridden by another explicit visibility assignment. This visibility of the content also applies to contents lower in the content hierarchy as well. Intuitively, {\em Nina} makes {\em NinaPhoto} visible to all, but makes it invisible for {\em Michiganders}. Hence, {\em Sue} will see her photo, but nothing else. On the other hand, {\em Prema} will see nothing although {\em Michiganders} is a subgroup of {\em all}, and by default should inherit the visibility right of her photo from {\em all}. This is because {\em Nina} made (essentially revoked the visibility right) {\em Everything} invisible to {\em Michiganders}, and {\em NinaPhoto} being a member of {\em Everything}, this invisibility applies to this object as well.

\subsubsection{Implied Visibility}
\label{implied-vis}

The implication of a set of explicit visibility assignments in our model is intricate and yet far more refined than any other known visibility models. Consider, for example, {\em Taylor}. {\em Nina} allowed him access to her personal information and hence, he will see everything under the node {\em PersonalInfo} in her content hierarchy. In addition, he will also see {\em Nina}'s {\em Blog} because she made it visible for {\em PistonFans}, and {\em Taylor} is a member of this group. However, {\em Bob} will only see {\em Nina}'s {\em Blog} and photo, and {\em JJ} will see everything. Notice that the visibility assignment of {\em Blog} and {\em NinaPhoto} for {\em PistonFans} overrides the visibility assignment of {\em Michiganders}. An implication of this reassignment is that the invisibility of all her personal information, except the photo, remain in effect for all members of {\em Michiganders}, except otherwise reassigned/overriden by her.

\subsubsection{Visibility Protocol}
\label{protocol}

Since we allow members to belong to multiple groups and arbitrary group hierarchies, it is possible that a member may inherit conflicting visibility rights via alternate membership hierarchies. For example, {\em JJ} inherits visibility right for {\em Nina}'s phone number by being a member of {\em UMichStudents} group, but is prohibited from accessing it by being a member of {\em Michiganders}. Perhaps {\em Nina} fears that since she allowed {\em PistonsFans} to see her photo, it is not safe to share her phone number with someone living near her home in Michigan who is also a Piston fan, while she trusts all UMich students. Now the question is should {\em JJ} be allowed access to {\em Nina}'s phone number? The answer lies in the visibility protocol adopted. We identify two protocols -- {\em optimistic} and {\em pessimistic}. {\em JJ} will see the phone number under the optimistic protocol where we accept the most permissive access right, and disallow it under a pessimistic protocol when conflicting rights are assigned, implicitly or otherwise. So, even under the optimistic protocol, {\em Bob} will only see her photo and {\em Blog}.

\subsubsection{Membership Reassignment}
\label{reassignment}

Grouping of members in a network is often accomplished by complex clustering algorithms, and such clusters are often recomputed as the population grows and their attributes change over time. Since content visibility to such large groups of members is usually assigned via group assignment rather than individual assignments, member visibility rights may change if their group assignment changes in a manner that affects their implied visibility rights. For example, consider members {\em Sue, Bob, Taylor} and {\em Mike}. Initially, there was no {\em Yankees} group, and {\em Sue} was not a member of any group. At a later time, a new group {\em Yankees} is created by the system administrator.  {\em Sue}, as well as {\em Bob} and {\em Taylor}'s membership has changed to this group shown using dashed lines, and the dashed {\em Yankees} group. A new member {\em Mike} is also added to this group. Now the question is, what visibility rights do each one of them have to {\em Nina}'s contents? In our model, {\em Mike} will inherit visibility right to {\em Nina}'s photo via the group {\em all}, and {\em Sue} will retain her right to {\em Nina}'s photo as well. However, {\em Bob} will lose his right to {\em Nina}'s {\em Blog} and {\em PistonPhotos} because access to those contents was through implicit rights via {\em PistonFans}, a group to which he no longer belongs. On the other hand, {\em Taylor} will continue to enjoy his access to {\em Nina}'s personal information due to the explicit right assignment, but lose his access to her {\em Blog}. In other words, implicit access right assignments are based on group memberships, and explicit access rights are not affected by group reassignment or reorganization, depending on the visibility protocol used. That means that even though explicit assignment is retained, conflicting rights may result in a more restrictive access in case of pessimistic protocol and more permissive access in an optimistic protocol.

\subsection{User Defined Groups}
\label{user-def-grp}

To make group assignments easier, we allow users to create their own group hierarchy consistent with our overall model. Recall that the network is modeled as a DAG and so we are allowed to have multiple roots. So the overall social network consists of the system defined network hierarchy and all the individual hierarchies defined by members. This feature helps members manage their visibility assignments better because they are now not required to speculate the implication of their visibility assignment in a large and complex network. The visibility protocol supplements this process by allowing the member to assign an optimistic protocol to her own network if she wants members in her network to retain the visibility rights she assigned even when the group reassignment takes effect at the system level. She chooses pessimistic protocol if she is uncertain or has doubts. For example, let us say she allows every one to see her phone number in her own network, but when a member in that network is grouped as a sex offender, she might not want that access to continue. Adoption of the pessimistic protocol will help such a default effect.

%\newpage

\section{Default Privacy Policies}
\label{default}

While a considerable argument persists over what should be private \cite{Marsoof11,GehrkeLP11,BrandtzaegLS10,Rosenblum07} and how privacy should be ensured \cite{KavianpourIM11,AhnSS11,TalukderOEEY10,FongAZ09,CarminatiFHKT11}, all agree that having a set of privacy rules is a core requirement. Recent research, however, suggest that complicated privacy policies usually deter users from using privacy settings optimally. Behavior and user interaction studies also reveal that typical users tend not to change their default privacy settings mainly due to the burden imposed in changing the defaults. On the other hand, in a complex network of millions of users, setting privacy rules is a formidable challenge for any casual user. These observations make a compelling argument in favor of a model for smart and effective default privacy rule generation \cite{FangKLT10,TochSH10}. There has been significant interest toward default privacy rule generation in general for information repositories, but such endeavors for social networks have been quite modest. We focus on two proposals that advocate default privacy rules -- one from the point of view of user interactions, and the other from a more technical standpoint.

\subsection{Current State}

Toch, Sadeh and Hong \cite{TochSH10} propose a collaborative privacy policy model in which they advocate a machine learning based suggestion of default rules that are flexible, configurable and highly likely to be accepted by the user. At the system level, new users are presented with a limited set of personalized default rules to choose from and configure. Default policies are learned feedback of select groups of experienced users of a location based social network system called Locaccino \cite{SadehHCFKPR09}\footnote{In a location based social network, geographical position is a prime attribute, e.g., location of a coffee shop and the users who are around it.}. In this model, each rule has three parts: manually selected groups of friends or peers who are allowed to see their location, the temporal window during which the location is shared, and finally locations that can be shared or disclosed. The learning algorithm generated rules such as 1) share location only when the user is on campus, 2) deny all requests and 3) share location at all times.

Danezis \cite{Danezis09}, on the other hand, argues that context of the information should be the prime driver for disclosure decision because it is a more appropriate indicator of relevance and perhaps ``the right to know" as well. Context can be dynamic and may change over time or space, and may render information accessible now but was prohibited before. From this conceptual point of view, the approach is substantially powerful and interesting. The approach focuses on supporting users wishing to restrict visibility to a subset of users of the information generated as part of interaction with other users. For such interactions, the model aims to infer the context boundary within which the interaction should reside and disallow users access to the interactions outside the boundary. They propose algorithms for inferring user context, context assignment to users,  and assignment of default visibility to the interaction.

\subsection{Strengths and Limitations}

There is a legitimate argument for default visibility rules and these two, and other similar proposals, make that argument convincingly from complementary viewpoints. An effective policy suggestion has the potential of actually encouraging users to use privacy settings beneficial for commerce and the community. In contrast, ineffective suggestions may lead to abuse and breach of privacy. Despite their strengths and potentials, the limitations are relatively more pronounced. For example, Locaccino's discovered rules are quite rudimentary and it is not clear if this model can actually be replicated in other types of social networks where more interesting and less obvious rules can be learned and suggested. Danezis' context discovery is based on techniques requiring substantial hints and inputs from users, and has the potential of discouraging rather than helping. Regardless of the technical limitations, default rules have fundamental non-user specific characteristics, making them less appealing to discriminative users. Therefore, if the suggested defaults do not have user specific features offered without much feedback or input from users, the chances of their effective use and acceptance remain doubtful.

\section{Dynamic Network Structures}
\label{dynamic}

The recent interest in exploiting social networks for commercial purposes \cite{KahlCTR11,AdnanNKTRR11,Chou10,ZhaoLLC09,LiBS11,YeungI11} has transformed them into a huge industry. One of the primary means for many commercial activities depend upon identifying and targeting users or groups. Network administrators often group users into a hierarchy or classification scheme. In many networks, users are often free to join in multiple such groups. As the interest, characteristics, and profile of users change, the network evolves. So the question arises, what happens to privacy settings of users in the structurally morphed network? While the answer to this question is still elusive, research in network reorganization is very active \cite{Schneier10a,BhagatCKS10,HeVSAA09,AbdessalemD11}.

\subsection{Current State}

Schneier \cite{Schneier10a} introduces the idea that not all data should have identical privacy rights. He classifies data into different classes and argues that some data should have revocable rights, and some may not. Derived information remains a murky category, i.e., should FaceBook be allowed to sell our browsing history for commercial purposes? This relates to the notion of user classification and network organization, and content ownership. The structure aware anonymization method proposed in \cite{HeVSAA09} is focused on protecting Schneier's derived behavioral data. The idea is to prevent identification of vital private user information by making it hard to do so when behavior networks are published by service providers for commercial means.  The question remains: How firmly does anonymization protect privacy when the underlying network evolves? Bhagat et al. \cite{BhagatCKS10} presents a method to quantify the degree to which privacy may be preserved in an evolving network by proper selection of anonymization properties.  A more subtle breach of privacy is by transitive disclosure \cite{AbdessalemD11} -- when users disclose information to a friend and another friend gains access to it from him. This research demonstrates that privacy guarantee improves when permissions are expressed in terms of reachability constraints over members of the network and user properties.

\subsection{Strengths and Limitations}
Most social networks essentially have two structures -- one that is visible to the public that composes users' communities, and their transitive relationships; the other is what the network administrators see and create through various analytical means that usually remain invisible. Most of the leading research addresses issues related to visible, and not so much the invisible, network. Anonymization and transitive disclosure research mainly concern visible networks and often do not take into consideration issues Schneier raises and content classification he proposes that have serious legal and social consequences. The major issue with the dynamic behavior of the invisible networks perhaps is their commercial implications. The jury is still out on what happens to invisible network super structures that site administrators create for commercial purposes or for social control such as law enforcement. How such use violate peoples' free speech rights, privacy and so on is still an open question and remain rarely discussed and addressed.

\section{User Controlled Privacy}
\label{user}

There are two complementary dimensions of the privacy issue -- users' responsibility to safeguard their information by specifying appropriate privacy rules, and the fiduciary responsibility of system administrators to protect user information (explicit or implicit). In this dual scheme, users need tools to express their privacy settings \cite{BadenBSBS09} and understand the implications of their rules in the evolving network \cite{FangL10,AnwarFYH09}. These rules must be robust and expressive enough to fend off any privacy breach using the visible network. Once expressed, the network administrator must ensure that the user rules are respected at all times \cite{HoadleyXLR10,YuanCY10} regardless of their interest in a third party relationship (e.g., using user information for commercial interests)  based on perhaps Schneier's content classification \cite{Schneier10a}. The recent incident involving Representative Anthony Wiener's Twitter messages and exposing deleted photos \cite{Danezis09} by the network exemplifies this duality. Had Twitter had ``better" privacy settings, perhaps Rep. Wiener could have avoided his career ending scandal (barring his moral behavior). On the other hand, when a user deletes a photo, the intent should be clear and allowing access to deleted content undoubtedly is a serious privacy violation \cite{Danezis09} and real loss of privacy rights as opposed to an illusory loss \cite{HoadleyXLR10}.

\subsection{Current State}

Persona \cite{BadenBSBS09} is a system that empowers users with cryptographic tools to control visibility of their contents. In this system, fine grained, attribute based public-private key based encryption is used to disclose information. Users are expected to manage these keys with a well defined distributed responsibility among peers and a network organization that groups users into categories. Once set up, the implications of the settings can be derived using research such as \cite{FangL10,AnwarFYH09} to understand the full privacy consequence. Using visualization tools, a user can see especially if an unwanted user or group is likely to gain access to the sensitive content he has made accessible to friends.

Yuan, Chen and Yu's \cite{YuanCY10} network publication scheme is along similar lines of research in network anonymization \cite{HeVSAA09} and denial of link discovery \cite{AbdessalemD11}. They, however, introduce the notion of personalization in the anonymization process and argue that user's privacy requirements should dictate how published networks are anonymized. They introduce the notion of attacker's background knowledge about users and the network, and a three-level protection scheme. Given a constant $k$ and a set of assumed attacker's background knowledge, the protection scheme guards against identification of a user (level 1), a relationship edge (level 2) and a relationship edge involving two specific users (level 3) with a probability $\frac{1}{k}$. The basic principle used is creating enough smokescreen to confuse the attacker. Higher $k$ ensures better safety, but increases the cost to ensure it too.

\subsection{Strengths and Limitations}

While these two methods are interesting in their own rights, they have substantial drawbacks. Technical limitations aside, Persona's passcode and key maintenance approach may be a headache that users are least likely to accept. User studies already show how reluctance in accepting a less than simple and ready to go set of policy settings \cite{TochSH10}. Given the growing participation in the digital world, users have so many virtual personalities that forgetting passwords is a frequent phenomena severe enough for sites to include password and identity retrieval help. Users are known to use the same password and reset hints in almost all their digital logins indirectly helping hackers crack them, creating a significant security risk \cite{TochSH10,TamGV10,YangC10,ShayKKLMBCC10,NovakovicMD09}. In addition to contributing to the encryption key management debacle, Persona's encryption scheme also suffers from huge maintenance overhead. For example, updating group membership discontinuation or network reorganization forces a rekeying of all keys in the subnetwork.

The purpose of publishing a network is perhaps for a specific and useful objective. Although the publication scheme presented in \cite{YuanCY10} deters attackers from finding information in the published networks, it is not clear how effective the publication scheme is to the stated objectives of the networks. It is also not clear how this scheme adjusts to the dynamic behavior of the network as it evolves.

\section{Access Control as a Privacy Theory}
\label{theory}

As we have alluded to earlier, sensitive social network information basically are of two types -- user information such as profiles and friends network, and derived information from the behavior and user profiles in the community they participate. Access control theory as a privacy instrument is mainly concerned with the protection of basic information. In recent years, several models have been proposed, and some have gained serious attention. Among them are the rule based privilege computation model \cite{CarminatiFHKT11}, privacy aware keyword search \cite{BjorklundGG10}, and access control based on ontology \cite{MasoumzadehJ10}, learning \cite{ShehabCTSC10}, evolution \cite{CrescenzoL09} and roles \cite{GulyasSI09}.

\subsection{Current State}

The basic model for access control in social network involves an information request or a query by a user. The response is computed by the network considering what data a user has access to where the space of accessible data is decided by user privacy settings. Since the network is primarily responsible for computing the response and ensuring privacy, machineries need to be in place to compute access right of the users, identify the accessible data space, and process the query efficiently.

The rule based privacy language of Carminati et al. \cite{CarminatiFHKT11} focuses on determining user access rights based on access control policies such as, {\em if Bob is a friend of Nina and Nina has a photo, Bob has read access to the photo.} Or a more restrictive rule is {\em if Bob is a friend of Nina, Nina has a photo and she granted access to Bob, Bob has read access to the photo.} The rule language is proposed in the context of semantic web based systems in which SWRL is used to encode security rules. Provisions are created to declare access control authorizations and prohibitions stating what privileges are allowed or disallowed. Then a reasoner is used to decide if access to a data item can be granted to a subject with a given privilege level.

Many search engines including Google and Bing support keyword search over Twitter posts, and Facebook allows searching of friends' posts. Often such posts and communications are private or restricted to friends, and friends of friends. The inverted indexing scheme proposed in \cite{BjorklundGG10} aims to efficiently compute search queries over such data considering privacy settings of users which tend to become inefficient due to a large number of redundant documents accessible by several users. They have shown that using their indices for posts and users, scalability can be improved and privacy preserved.

\subsection{Strengths and Limitations}

One of the major advantages of these models is that they empower the user and enhance the security apparatus that protects basic information. If the protection of basic information is guaranteed, privacy models for derived information will also benefit. The implication is that this will offer clarity about public view of user contents and we will then be left with deciding what derived information are legally public or usable without user consent. The space of this information is expected to be smaller and less complex than it is now.

The limitations, however, in general are pronounced. For example, though unique in their approach, the search technique introduced in \cite{BjorklundGG10} does not consider the actual network which is always evolving. It is not known how the ranking function the authors proposed would behave in real networks. The rule based model \cite{CarminatiFHKT11} though exciting, relies heavily on system level privacy policies that may be too general for users who tend to avoid using generic privacy settings or use it improperly \cite{TochSH10,TamGV10,YangC10,ShayKKLMBCC10,NovakovicMD09}. This language also has limited primitives to support user privacy settings, preferences, and community building and participation. The rule specification language in ontology assisted privacy model proposed in \cite{MasoumzadehJ10} is complex enough to serve as a bottleneck. While the classifier based learning approach in \cite{ShehabCTSC10} offers control to users, the management and the complexity associated with the selection of the training data, parameter setting and evolution of the network may act as serious deterrents. Despite their technical limitations, research in this direction seems to be useful and appealing.

\section{Query Language for Privacy Enforcement}
\label{language}

Social networks being a huge database, naturally opportunities exist for designing languages to query and analyze their contents \cite{CarminatiFP06,CarminatiFHKT11,RonenS09,MartinGW11,DriesNR09}. Traditionally in databases, authorization and access control has been left as an orthogonal issue from query languages amounting to view generation as an all or nothing proposition, although fine grained authorization has been investigated, e.g., in the context of multi-level security \cite{JajodiaS91}. Social networks being graph structured data, query languages for graphs also become relevant \cite{GraphQL}. The distinction in the case of social network data is that each user data privacy policy is distinct from the rest of the users, and thus addressing access control in bulk as in relational model seems unworkable, or at least inefficient. A fine grained method instead appears more appropriate. Given the complexity of the data type, representation and granularity, making privacy provisions at the query language level may yield a better platform.

\subsection{Current State}

SNQL \cite{MartinGW11} is a pattern query language for social network databases that has provisions for creating new network structures and analysis. It follows on the footsteps of a long chain of research in pattern query languages such as BiQL \cite{DriesNR09}, SocialScope \cite{Amer-YahiaLY09} and SoQL \cite{RonenS09}. Though not specifically designed for social networks, GraphQL is a declarative pattern language designed for network data querying with complementary features for structures. The {\sf construct where from} structure has a flavor similar to SQL's {\sf select from where}. Expressive but complicated triple patterns and conditions may be expressed respectively in the {\sf construct} and {\sf where} clauses. The rule language of Carminati et al. \cite{CarminatiFP06,CarminatiFHKT11} is another prominent example of this category of systems that uses query language to support privacy (see previous section for more discussion on this language).

\subsection{Strengths and Limitations}

These two complementary class of languages has obvious strengths, but they also suffer from critical shortcomings. In particular, the rule language is not designed to support user specification of privacy rules. Instead a set of rules is treated as privacy axioms that apply to all. Similar to SNQL, it is also unaware of the evolving nature of the network. More importantly, it does not particularly consider user groups, hierarchies and exceptions of privacy rules that may apply. But the idea that rules can be applied to decide privileges is an interesting concept not found in many approaches. SNQL on the other hand, is focused on structure manipulation and treats privacy as a separate issue.

\section{Outline of a Privacy Query Language\\ for Social Networks}
\label{outline}

In view of the issues raised in sections \ref{default} through \ref{language}, and the reference conceptual model for social networks introduced in section \ref{model}, we now discuss a possible query language that can address most of them directly and help address the remaining ones indirectly. We identify features and concepts that we can assemble in a language we believe will be sufficient to address them. Our position is that once such a language is available, it can be extended with network and pattern analysis capabilities to incorporate features available in SNQL \cite{MartinGW11}, SocialScope \cite{Amer-YahiaLY09} and SoQL \cite{RonenS09}.

We believe many of the disclosure related problems in social networks could be remedied by adopting a content classification scheme similar to Schneier \cite{Schneier10a} that will be able to withstand any legal scrutiny, and then designing a query language that respects the disclosure rules implied by this classification. Since legality is a time varying concept, the language will need to be sensitive to this as well. Once a query language is available to access legally disclosable information, secondary information contents can be derived from this accessible set. Only issue we then will have to deal with is how we share and use information that are not legally public, which we believe is arguably a smaller set. In our vision, a query language that allows some form of reasoning, perhaps a rule-based framework, with the capability of inheritance with overriding in class hierarchies stand the best chance of addressing most of the issues we raised in sections \ref{default} through \ref{language} in a comprehensive manner in the context of the reference model introduced in section \ref{model}.

For example, many object-oriented query languages support objects, class hierarchies, inheritance, overriding, encapsulation and signatures. The class hierarchy feature may be used to organize members and groups in the manner described in sections \ref{groups} and \ref{hierarchy} as part of our reference model. Members in a group can be made to adhere to specific structures such as having specific profiles and contents as in many object-oriented databases (mandatory and optional properties). An auxiliary system may use these properties, along with other properties, to classify and reclassify members as needed. Users may also create member groups for their own purposes as discussed in section \ref{user-def-grp} to gain greater control in assigning access rights.

Visibility of properties and contents (discussed in sections \ref{content-vis} and \ref{implied-vis}) owned by members may be captured as signatures in a way similar to the notion of encapsulation in object-oriented databases to control access to protected contents. Then, as opposed to inheritance of default class values and methods, the visibility rules captured as signatures may be allowed to be inherited in the class hierarchy with overriding to model fine-grain visibility as definition specificity. By adopting a ``choice mechanism" among a set of visibility rules, we could implement the notion of optimistic and cautious visibility protocols introduced in section \ref{protocol}. Such a choice mechanism will help deal with conflicts in multiple inheritance hierarchies and class structure reorganization in fairly simple ways.

Since we allow class reorganization, conflicting access rights are likely to result due to membership in multiple groups even though overriding of access rights in the inheritance hierarchy will ensure singular visibility in a specific inheritance path. Parameterization of traditional encapsulation may be used to grant access to specific groups or members, as opposed to global access to everyone often found in object-oriented systems. Further parameterization to include cautious and optimistic protocols may be used to allow a more finer grained access and privacy. For example, {\em Nina} may allow all {\em PistonFans} to see her {\em age} in cautious mode, but her {\em NinaPhoto} picture in optimistic mode. So, in the event of a conflict, {\em PistonFans} may not see her {\em age}, but they will always see her picture if she granted read permission to both. Not that, to allow uninterrupted access (or no access) to anyone or a group, all a member has to do is specify access rule for that group or member specifically and no reorganization will impact that declared visibility as this declaration will override all other declaration. Recall that conflicts arise only due to no specific rules and a group or member inherits conflicting visibilities.

The language we outlined above has already addressed the issue of dynamic structures (discussed in \ref{dynamic}), user controlled privacy (discussed in section \ref{user}), and query language (discussed in section \ref{language}). The issue of default privacy policy discussed in section \ref{default} is partially addressed in the proposed language, i.e., by default all contents and properties are invisible. But tools can be developed to assist users to determine what default privacy settings are useful and safe for them on a case by case basis, and many of the systems discussed may be adapted to serve this need. Since we advocate a rule based language, deductive privacy settings discussed in section \ref{theory} can now be easily implemented using visibility rules at the desired granularity. We foresee no need for any specific tool or separate mechanism to address this issue.

\section{Conclusions}
\label{conclusion}

The focus of this paper was to briefly survey contemporary research on privacy in social networks in four complementary axes. We have proposed a conceptual model social network style for digital communities with privacy. In this model, users are given the tools and the prime responsibility of defining their own privacy rules with a simple semantics. We believe that the proposed model fares well with these criteria. As presented in sections \ref{default} through \ref{language}, most models favor one or the other feature and are not comprehensive enough. We believe it is because the model used in each research was not designed anew from the ground up, and thus retained the original shortcomings. The discussion in each of these sections summarize why we believe these systems need to accommodate the features proposed in our model.

The distinctive features of our model include an explicit mechanism for users to define a DAG like structure for grouping communities, and allowing the system to do the same for its own purposes. We have shown that hierarchy of users and objects make it possible to model privacy and exceptions using non-monotonic inheritance of visibility privileges. The interesting property is that network reorganization does not require any corrective measure to reestablish user view of privacy. Since privacy is modeled as visibility rules at the user level on the basic view of the network, it is expected that any derived view will not violate the privacy view of users. We, however, used only one class of content and a binary visibility setting, although Schneier's \cite{Schneier10a} view of content classification can also be accommodated in this model.

In \cite{PiQL-PODS-2012}, we have already proposed a declarative privacy query language based on the reference model introduced in this paper. In \cite{PiQL-PODS-2012}, we have demonstrated that the features of the model introduced in this paper can actually be cast into a sound and complete first-order object-oriented language. We are currently developing a visualization toolkit to assist users grasp the implications of their privacy specifications, and offer access to content through queries. While we did not include a discussion on an explicit privacy implication toolkit in this paper, this can be added orthogonally along the lines of Fang and LeFevre \cite{FangL10} and Anwar et al. \cite{AnwarFYH09} using which, users will be able to comprehend the implications of their privacy settings. It is also possible to develop default setting suggestions tool kits along the lines of \cite{SadehHCFKPR09,Danezis09}.

%\newpage

\bibliographystyle{IEEEtran}
%\bibliography{/bib-db/bib-db-general,/bib-db/our-publications}


\end{document}